\documentclass[submission,copyright,creativecommons]{eptcs}
\providecommand{\event}{GraMSec 2014 -- First International Workshop on Graphical Models for Security} 
\usepackage{breakurl}             

\usepackage{graphicx}

\usepackage{bussproofs}

\usepackage{amssymb}
\usepackage{amsmath}

\usepackage{hyperref}

\newtheorem{theorem}{Theorem}[section]

\newtheorem{example}[theorem]{Example}

\newcommand\confNames{\ensuremath{\mathcal{N}_{C}}}
\newcommand\chanNames{\ensuremath{\mathcal{N}_{A}}}
\newcommand\identNames{\ensuremath{\mathcal{N}_{I}}}
\newcommand\psiTerms{\ensuremath{\mathbf{T}}}
\newcommand\psiConditions{\ensuremath{\mathbf{C}}}
\newcommand\psiAssertions{\ensuremath{\mathbf{A}}}
\newcommand\psiChanEq{\ensuremath{\stackrel{\ldotp}{\leftrightarrow}}}
\newcommand\psiCompAssert{\ensuremath{\otimes}}
\newcommand\psiAssertUnit{\ensuremath{\mathbf{1}}}
\newcommand\psiEntailment{\ensuremath{\vdash}}
\newcommand\psiassertion{\ensuremath{\Psi}}
\newcommand\psiEquivAssert{\ensuremath{\simeq}}
\newcommand\psiEmptyProc{\ensuremath{\mathbf{0}}}
\newcommand\psiOutput[2]{\ensuremath{\overline{#1}\langle#2\rangle}}
\newcommand\psiInput[2]{\ensuremath{\underline{#1}\langle#2\rangle}}
\newcommand\psiAssertOp[1]{\ensuremath{(\hspace{-0.07cm}\lvert#1\rvert\hspace{-0.07cm})}}
\newcommand\psiCase[1]{\ensuremath{\mathbf{case}\ #1\,}}
\newcommand\psiNew[1]{\ensuremath{(\nu #1)}}
\newcommand\psiframe[1]{\ensuremath{\mathcal{F}(#1)}}
\newcommand\psiPar{\ensuremath{\,|\,}}
\newcommand\psiRec{\ensuremath{!}}
\newcommand\internalAction{\ensuremath{\tau}}
\newcommand\dependsPomset{\ensuremath{\leq}}
\newcommand\psitransition[4]{\ensuremath{#1\vartriangleright #2 \xrightarrow{#3} #4}}

\newcommand\defequal{\ensuremath{\stackrel{\vartriangle}{=}}}

\title{Actor Network Procedures as Psi-calculi\\ for Security Ceremonies
\thanks{This work was partially supported by the project \href{https://www.offpad.org/}{OffPAD} with number E!8324 part of the \href{http://www.eurekanetwork.org/activities/eurostars}{Eurostars} program funded by the \href{http://www.eurekanetwork.org}{EUREKA} and European Community.}
}
\author{Cristian Prisacariu
\thanks{\textbf{Acknowledgements:} I would like to thank Audun J\o{}sang for introducing me to security ceremonies and for explaining their practical usefulness and the need for an adequate formalism (i.e., graphical, expressive, and with reasoning capabilities).}
\institute{Dept. of Informatics, University of Oslo, \ -- \ P.O.\ Box 1080 Blindern, N-0316 Oslo, Norway.}
\email{cristi@ifi.uio.no}
}
\def\titlerunning{Actor Network Procedures as Psi-Calculi for Security Ceremonies}
\def\authorrunning{C.~Prisacariu
}
\begin{document}
\maketitle

\begin{abstract}
The \textit{actor network procedures} of Pavlovic and Meadows are a recent graphical formalism developed for describing security ceremonies and for reasoning about their security properties. The present work studies the relations of the actor network procedures (ANP) to the recent psi-calculi framework. 
Psi-calculi is a parametric formalism where calculi like spi- or applied-pi are found as instances. Psi-calculi are operational and largely non-graphical, but have strong foundation based on the theory of nominal sets and process algebras.
One purpose of the present work is to give a semantics to ANP through psi-calculi. Another aim was to give a graphical language for a psi-calculus instance for security ceremonies. At the same time, this work provides more insight into the details of the ANPs formalization and the graphical representation.
\end{abstract}

\section{Introduction}

\textit{Actor Network Procedures} (ANP) is a formalism introduced in \cite{pavlovic11ActorNet_journal} for modeling security ceremonies \cite{ellison07ceremony,RadkeBNB11ceremony}. Reasoning about security properties of ceremonies is done using the Procedure Derivation Logic, which comes from a line of research on logics for composition of protocols that started with the Protocol Composition Logic \cite{DurginMP03PCL,mitchell11PCL}. This formalism that we concentrate on in this paper should not be confused with the work with similar purposes from \cite{pieters2011representing}. Both these approaches \cite{pieters2011representing,pavlovic11ActorNet_journal} aim to be used for describing security ceremonies by drawing inspiration from the actor network theory in sociology, where the book \cite{latour05book} gives a good overview.

Security ceremonies are well motivated in \cite{ellison07ceremony,RadkeBNB11ceremony} with convincing examples. Technically, security ceremonies are meant to extend security protocols by including the human in the formalization and making explicit the environment (and the attacker). A ceremony may also combine protocols. In consequence, a formalism for security ceremonies is expected to be expressive enough to include existing formalisms for protocols as special cases. Such a formalism should offer the possibility to model human behavior related to the ceremony. Since ceremonies would tend to be large, because of all the assumptions that are included explicitly, we expect compositionality to be a main aspect of the formalism, both for design and for reasoning.

For usability purposes a desired formalism for security ceremonies would offer a graphical language for developing the ceremonies, as well as for reasoning. Yet the graphics should be formally grounded, so to have guarantees for the security results obtained. A good example of such formally grounded graphical languages can be the statecharts \cite{Harel87statecharts,HarelN96statemate} or the live sequence charts \cite{harel03Play_book}, which were intended for describing concurrent and reactive systems.

The aim of the actor network procedures \cite{pavlovic11ActorNet_journal} is to be graphical, expressive, compositional, and with formal logical reasoning capabilities. In this paper we look more carefully at this formalism and relate it to the psi-calculi framework, which offers solid semantical analysis and possibility for correlations with existing security formalisms coming from the process algebra approach.

Psi-calculi \cite{11psi_journal} are a semantics framework where various existing calculi can be found as instances. In particular, the spi- and applied-pi calculi \cite{AbadiG99spi_calculus,AbadiF01applied_pi} are two instances of interest for security protocols. Psi-calculi are though a non-graphical algebraic formalism; but with strong mathematical background. Results and tools of psi-calculi, e.g., involving theorem provers or true concurrency semantics, could be thus used in the setting of actor network procedures. Nevertheless these could easily be hidden from the security ceremony developer behind the graphical formalism ANPs provide.

With the danger of seaming somewhat pedantic to some readers, a little bit more motivation for use of formal techniques for security protocols is in order here. Arguably, for security systems perfection and assurance of perfection are highly important, since bugs cannot be considered ``features'', as is the case in other areas. For a security system one often wants to be provided with guarantees that some expected security properties are met. This can be even more difficult to achieve for security ceremonies, which are more complex, composing protocols, including hidden assumptions and human models. 
In the case of security protocols one hardly can rely on testing to provide assurance; and experience has shown that protocols that are thoroughly tested in practice for years turn out to contain serious flaws, where a famous example is \cite{Lowe96}.

This motivates why considerable amounts of research have been put into providing mathematical models and theories for studying security protocols. But more practical are the formal tools that have been built on top of these theories so to have a (semi-)automated way of ensuring security properties. Examples of tools include model checkers like Murphi \cite{MitchellMS97murphi,SternD97murphi_paralel} or FDR \cite{fdr_book} which are push-button tools with yes/no answers; or theorem provers like ProVerif \cite{Blanchet04proverif} for the symbolic (process calculi) approach and CryptoVerif \cite{Blanchet08cryptoverif} for the computational approach, or Isabell/HOL \cite{Paulson98isabell}, which often need interaction with expect users but which achieve stronger results than model checkers do.

The psi-calculi is a framework that goes well with the ProVerif and FDR tools which are also based on process calculi. But more than this, psi-calculi have been built (i.e., all the related meta-results have been proven) using the proof assistant Isabell/HOL \cite{NipkowPW02}. Therefore one could say that psi-calculi could be seen as lying at the intersection of the two kinds of tools, making use of the strengths of both.

A downside of such strong formally grounded frameworks is that they are difficult to use by common developers of security protocols and ceremonies. This is where graphical formalisms usually can provide considerable simplifications and hide formal notation, concentrating on the concepts and methods instead. A quite appreciated example of graphical languages grounded in theory can be found in the area of developing reactive or concurrent systems. Here the groundbreaking was achieved through \textit{statecharts} \cite{Harel87statecharts} which has become a standard and which have been extended into the more recent \textit{live sequence charts} \cite{harel03Play_book}.
A long term goal of the present author is to have a similar graphical language and tool-set for security ceremonies; and this paper tries to identify a path in this direction.

\section{Background on Psi-Calculi}

\textit{Psi-calculus} \cite{11psi_journal} has been developed as a framework for defining nominal process calculi, like the many variants of the pi-calculus \cite{milner92picalcul}. The psi-calculi framework is based on nominal datatypes, and \cite[Sec.2.1]{11psi_journal} gives a sufficient introduction to nominal sets used in psi-calculi. We will not refer much to nominal datatype i this paper, but refer the reader to the book \cite{pitts_book_nominal} which contains a thorough treatment of both the theory behind nominal sets as well as various applications (e.g., see \cite[Ch.8]{pitts_book_nominal} for nominal algebraic datatypes). We expect, though, some familiarity with notions of algebraic datatypes and term algebras. 

The psi-calculi framework is parametric; instantiating the parameters accordingly, one obtains an \textit{instance of psi-calculi}, like the pi-calculus, or the cryptographic spi-calculus.
These parameters are:
\begin{center}
\begin{tabular}{ll}
\psiTerms & terms (data/channels)\\
\psiConditions & conditions\\
\psiAssertions & assertions
\end{tabular} 
\end{center}
which are nominal datatypes not necessarily disjoint; together with the following operators:
\begin{center}
\begin{tabular}{ll}
$\psiChanEq\ :\ \psiTerms\times\psiTerms\rightarrow\psiConditions$ & channel equality\\
$\psiCompAssert\ :\ \psiAssertions\times\psiAssertions\rightarrow\psiAssertions$ & composition of assertions\\
$\psiAssertUnit\ \in\ \psiAssertions$ & minimal assertion\\
$\psiEntailment\ \subseteq\ \psiAssertions\times\psiConditions$ & entailment relation
\end{tabular} 
\end{center}

Intuitively, terms can be seen as generated from a signature, as in term algebras; the conditions and assertions can be those from first-order logic; the minimal assertion being top/true, entailment the one from first-order logic, and composition taken as conjunction. We will shortly exemplify how pi-calculus is instantiated in this framework.
The operators are usually written infix, i.e.: $M\psiChanEq N$, $\psiassertion \psiCompAssert\psiassertion'$, $\psiassertion\psiEntailment\varphi$.

The above operators need to obey some natural requirements, when instantiated. Channel equality must be symmetric and transitive. The composition of assertions must be associative, commutative, and have \psiAssertUnit\ as unit; moreover, composition must preserve equality of assertions, where two assertions are considered equal iff they entail the same conditions (i.e., for $\psiassertion,\psiassertion'\in\psiAssertions$ we define the equality  $\psiassertion\psiEquivAssert\psiassertion'$ iff $\forall\varphi\in\psiConditions:\psiassertion\psiEntailment\varphi \Leftrightarrow \psiassertion'\psiEntailment\varphi$).

The intuition is that assertions will be used to assert about the environment of the processes. Conditions will be used as guards for guarded (non-deterministic) choices, and are to be tested against the assertion of the environment for entailment. Terms are used to represent complex data communicated through channels, but will also be used to define the channels themselves, which can thus be more than just mere names, as in pi-calculus. The composition of assertions should capture the notion of combining assumptions from several components of the environment.

The syntax for building psi-process is the following (psi-processes are denoted by the $P,Q,\dots$; terms from \psiTerms\ by $M,N,\dots$; ):
\begin{center}
\begin{tabular}{ll}
$\psiEmptyProc$ & Empty/trivial process\\
$\psiOutput{M}{N}.P$ & Output\\
$\psiInput{M}{(\lambda\tilde{x})N}.P$ & Input\\
$\psiCase{\varphi_{1}:P_{1},\dots,\varphi_{n}:P_{n}}$ & Conditional (non-deterministic) choice\\
$\psiNew{a}P$ & Restriction of names $a$ inside processes $P$\\
$P\psiPar Q$ & Parallel composition\\
$\psiRec P$ & Replication\\
$\psiAssertOp{\psiassertion}$ & Assertions
\end{tabular} 
\end{center}
The empty process has the same behavior as, and thus can be modeled by, the trivial assignment $\psiAssertOp{\psiAssertUnit}$.

The input and output processes are as in pi-calculus only that the channel objects $M$ can be arbitrary terms. In the input process the object $(\lambda\tilde{x})N$ is a pattern with the variables $\tilde{x}$ bound in $N$ as well as in the continuation process $P$. Intuitively, any term message received on $M$ must match the pattern $N$ for some substitution of the variables $\tilde{x}$. The same substitution is used to substitute these variables in $P$ after a successful match.
The traditional pi-calculus input $a(x).P$ would be modeled in psi-calculi as $\psiInput{a}{(\lambda x)x}.P$, where the simple names $a$ are the only terms allowed.

The case process behaves like one of the $P_{i}$ for which the condition $\varphi_{i}$ is entailed by the current environment assumption, as defined by the notion of \textit{frame} which we preset later. This notion of frame is familiar from the applied pi-calculus, where it was introduced with the purpose of capturing static information about the environment (or seen in reverse, the frame is the static information that the current process exposes to the environment).
A particular use of case is as $\psiCase{\varphi:P}$ which can be read as $\mathbf{if}\ \varphi\ \mathbf{then}\ P$. Another special usage of case is as $\psiCase{\top:P_{1},\,\top:P_{2}}$, where $\psiassertion\psiEntailment\top$ is a special condition that is entailed by any assertion, like $a\psiChanEq a$; this use is mimicking the pi-calculus nondeterministic choice $P_{1} + P_{2}$.
Restriction, parallel, and replication are the standard constructs of pi-calculus.

Assertions $\psiAssertOp{\psiassertion}$ can float freely in a process (i.e., be put in parallel) describing assumptions about the environment. Otherwise, assertions can appear at the end of a sequence of input/output actions, i.e., these are the guarantees that a process provides after it finishes (on the same lines as in assume/guarantee reasoning about programs). Assertions are somehow similar to the active substitutions of the applied pi-calculus, only that assertions do not have computational behavior, but only restrict the behavior of the other constructs by providing their assumptions about the environment.

\begin{example}[pi-calculus as an instance]\label{ex_pi_instance}
To obtain pi-calculus \cite{milner92picalcul} as an instance of psi-calculus use the following, built over a single set of names $\mathcal{N}$:
\begin{center}
\begin{tabular}{lcl}
\psiTerms & \defequal & $\mathcal{N}$\\
\psiConditions & \defequal & $\{a=a \mid a,b\in\psiTerms\}$\\
\psiAssertions & \defequal & $\{\psiAssertUnit\}$\\
$\psiChanEq$ &  \defequal & $=$\\
$\psiEntailment$ &  \defequal & $\{(\psiAssertUnit,a=a) \mid a\in \psiTerms\}$
\end{tabular} 
\end{center}
with the trivial definition for the composition operation.
The only terms are the channel names $a\in\mathcal{N}$, and there is no other assertion than the unit. The conditions are equality tests for channel names, where the only successful tests are those where the names are equal. Hence, channel comparison is defined as just name equality.
\end{example}

Psi-calculus is given an operational semantics in \cite{11psi_journal} using labeled transition systems, where the nodes are the process terms and the transitions represent one reduction step, labeled with the action that the process executes. The actions, generally denoted by $\alpha,\beta$, represent respectively the input and output constructions, as well as $\internalAction$ the internal synchronization/communication action: 
%

\vspace{1ex}\centerline{$\psiOutput{M}{(\nu\tilde{a})N} \mid \psiInput{M}{N} \mid \internalAction$}


Transitions are done in a context, which is represented as an assertion \psiassertion, capturing assumptions about the environment:

\centerline{$\psitransition{\psiassertion}{P}{\alpha}{P'}$}

\noindent Intuitively, the above transition could be read as: The process $P$ can perform an action $\alpha$ in an environment respecting the assumptions in $\psiassertion$, after which it would behave like the process $P'$.

The context assertion is obtained using the notion of \textit{frame} which essentially collects (using the composition operation) the outer-most assertions of a process.
The frame $\psiframe{P}$ is defined inductively on the structure of the process as:
\begin{center}
\begin{tabular}{l}
$\psiframe{\psiAssertOp{\psiassertion}} = \psiassertion$\\
$\psiframe{P\psiPar Q} = \psiframe{P} \psiCompAssert \psiframe{Q}$\\
$\psiframe{\psiNew{a}{P}} = \psiNew{a}{\psiframe{P}}$\\
$\psiframe{\psiRec P} = \psiframe{\psiCase{\tilde{\varphi}:\tilde{P}}} = \psiframe{\psiOutput{M}{N}.P} = \psiframe{\psiInput{M}{(\lambda\tilde{x})N}.P} = \psiAssertUnit$
\end{tabular} 
\end{center}
Any assertion that occurs under an action prefix or a condition is not visible in the frame.

We give only an exemplification of the transition rules for psi-calculus, and refer to \cite[Table 1]{11psi_journal} for the full definition. The \textsc{(case)} rule shows how the conditions are tested against the context assertions. The communication rule \textsc{(com)} shows how the environment processes executing in parallel contribute their top-most assertions to make the new context assertion for the input-output action of the other parallel processes.
In the (com)-rule the assertions $\psiassertion_{P}$ and $\psiassertion_{Q}$ come from the frames of $\psiframe{P}=\psiNew{\tilde{b_{P}}}{\psiassertion_{P}}$ respectively $\psiframe{Q}=\psiNew{\tilde{b_{Q}}}{\psiassertion_{Q}}$.
\begin{prooftree}
\AxiomC{$\psitransition{\psiassertion}{P_{i}}{\alpha}{P'}$}
\AxiomC{$\psiassertion\psiEntailment\varphi_{i}$}
\RightLabel{\textsc{(case)}}
\BinaryInfC{$\psitransition{\psiassertion}{\psiCase{\tilde{\varphi}:\tilde{P}}}{\alpha}{P'}$}
\end{prooftree}
\begin{prooftree}
\AxiomC{$\psitransition{\psiassertion_{Q}\psiCompAssert\psiassertion}{P}{\psiOutput{M}{(\nu\tilde{a})N}}{P'}$}
\AxiomC{$\psitransition{\psiassertion_{P}\psiCompAssert\psiassertion}{Q}{\psiInput{K}{N}}{Q'}$}
\AxiomC{$\psiassertion_{Q}\psiCompAssert\psiassertion_{P}\psiCompAssert\psiassertion\psiEntailment M\psiChanEq K$}
\RightLabel{\textsc{(com)}}
\TrinaryInfC{$\psitransition{\psiassertion}{P\psiPar Q}{\internalAction}{\psiNew{\tilde{a}}{(P'\psiPar Q')}}$}
\end{prooftree}

There is no transition rule for the assertion process; this is only used in constructing frames. Once an assertion process is reached, the computation stops, and this assertion remains floating among the other parallel processes and will be composed part of the frames, when necessary, like in the case of the communication rule.

\section{A Psi-Calculus Instance for Actor Network Procedures}

We do not introduce the notation and definitions used in the ANPs of \cite{pavlovic11ActorNet_journal} because our aim here is to develop teh ANP ideas in the formal language of psi-calculi. In consequence, we lie to see this section as a formal description of ANPs. The ideas and development from \cite{pavlovic11ActorNet_journal} of the ANPs require quite expressive theories which cannot be easily captured with traditional formalisms for security protocols, but which are available in the psi-calculi framework.

There are a few aspects of psi-calculi that offer us the possibility to give semantics to the actor network procedures in this section; and in fact to complex systems like ubiquitous systems where humans are part of the system.

One aspect is the expressiveness of the terms that are allowed to be used for representing messages. This is very liberal in psi-calculi, and thus can easily capture complex structures of messages. Moreover, nominals are allowed in the terms for capturing the important notion of names (like in pi-calculi, or object-oriented languages). Names appear in actor network procedures in three places, as we see further, as names for channels, system configurations, and names of participants in the ceremony.

Another aspect of psi-calculi is the way communication can be done through arbitrarily complex communication terms. This means we are not restricted to just one channel name, but more structure for channels is allowed. We exploit this when modeling the structure of the system configurations and their attached channels and participants. This more complex structure is responsible for the communications in the actor network procedures.

An important aspect of psi-calculi is also the logic that is available through the assertions and the conditions language, and the entailment relation between the two. The liberty that the psi-calculi framework offers for defining the logical part of the calculus allows one to choose the right language for the application purpose. In consequence, depending on the problem one can choose a more expressive logical entailment or one with better computation properties (i.e., feasible for automation).
One way of using the assertions and conditions is as in the Hoare-style of reasoning. We may have pre-conditions (using the case construct) and post-conditions (using the trailing assertions) for individual actions as well as for complex processes. Essentially, psi-calculi allow us to have an assume/guarantee reasoning style using the logical language of our choice and the granularity of the reasoning (i.e., process/action level). Processes are annotated with logical assertions so that an external logical reasoning system can be used on top of the process. But as well, the conditions are logically tested by the process while running, constituting a runtime level reasoning system.

The logical part of the psi-calculi will be exploited to capture the reasoning aspects that actor networks procedures and their PDL logic offers.

\subsection{Example of an Actor Network Procedure}\label{subsec_example}

We take the \textit{two-factor authentication} example from \cite{pavlovic11ActorNet_journal} of the Chip Authentication Program (CAP) procedure. The configuration structure is described in \cite[2.4.2]{pavlovic11ActorNet_journal} whereas the runs of the ceremony are described in \cite[3.5.1]{pavlovic11ActorNet_journal}. The graphical formalization of this ceremony is given in \cite[Fig.3]{pavlovic11ActorNet_journal} for the structure and in \cite[Fig.7]{pavlovic11ActorNet_journal} for the run; we will use the same original figures so to stick with the graphical choices of the authors of \cite{pavlovic11ActorNet_journal}. 

\begin{figure}[tp]
\begin{center}
    \includegraphics[height=4.1cm]{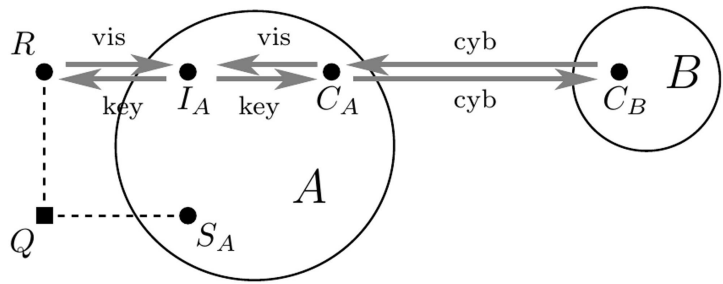}
  \end{center}
\vspace{-3ex}
\caption{The structure of the configurations, part of the CAP procedure.}
\label{fig_ex_CAP_structure}
\vspace{-2ex}
\end{figure}

The Figure~\ref{fig_ex_CAP_structure} (taken from \cite[Fig.3]{pavlovic11ActorNet_journal}) graphically represents the structure of the configurations and the attached communication channels for the actor network procedure that formalizes the CAP two-factor authentication \cite{DrimerMA09CAP}.
The structure contains two identity names $A$ (for Alice) and $B$ (for the Bank) which are attached (as subscript) to some of the configurations. The circled areas are not strictly necessary, and become impossible to represent for larger examples; but are good visual aid for examples like this one where they encircle all those configurations corresponding to the respective identity.

The single configuration $C_{B}$ represents the Bank's computer. Three configurations are under $A$'s control: the computer $C_{A}$, the card $S_{A}$, and the human representation of Alice $I_{A}$. A card reader $R$ is also available, which when coupled with Alice's card form the configuration $Q$. The human $I_{A}$ can output through a keyboard channel information to Alice's computer and through another keyboard channel to the card reader. Both the card reader and the computer have visual displays that give information to the human.
There are two cyber channels between the two computers; cyber channels are assumed to be untrusted and the information on them should be transmitted encrypted.

The arrows represent channels, and have attached a label denoting the type of the channel. The dark circles and squares represent configurations, where the squares are complex ones containing sub-configurations, whereas the circles are minimal configurations; which are called nodes in \cite{pavlovic11ActorNet_journal}. The dashed lines display the containment relation between the configurations.

\begin{figure}[tp]
\begin{center}
    \hspace{-5ex}\includegraphics[width=11.6cm]{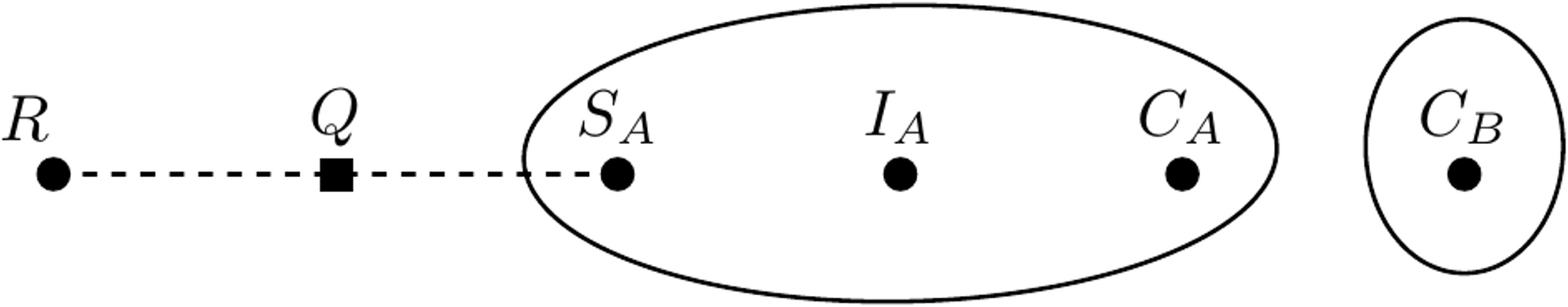}\\
    \vspace{-1ex}\includegraphics[width=12.5cm]{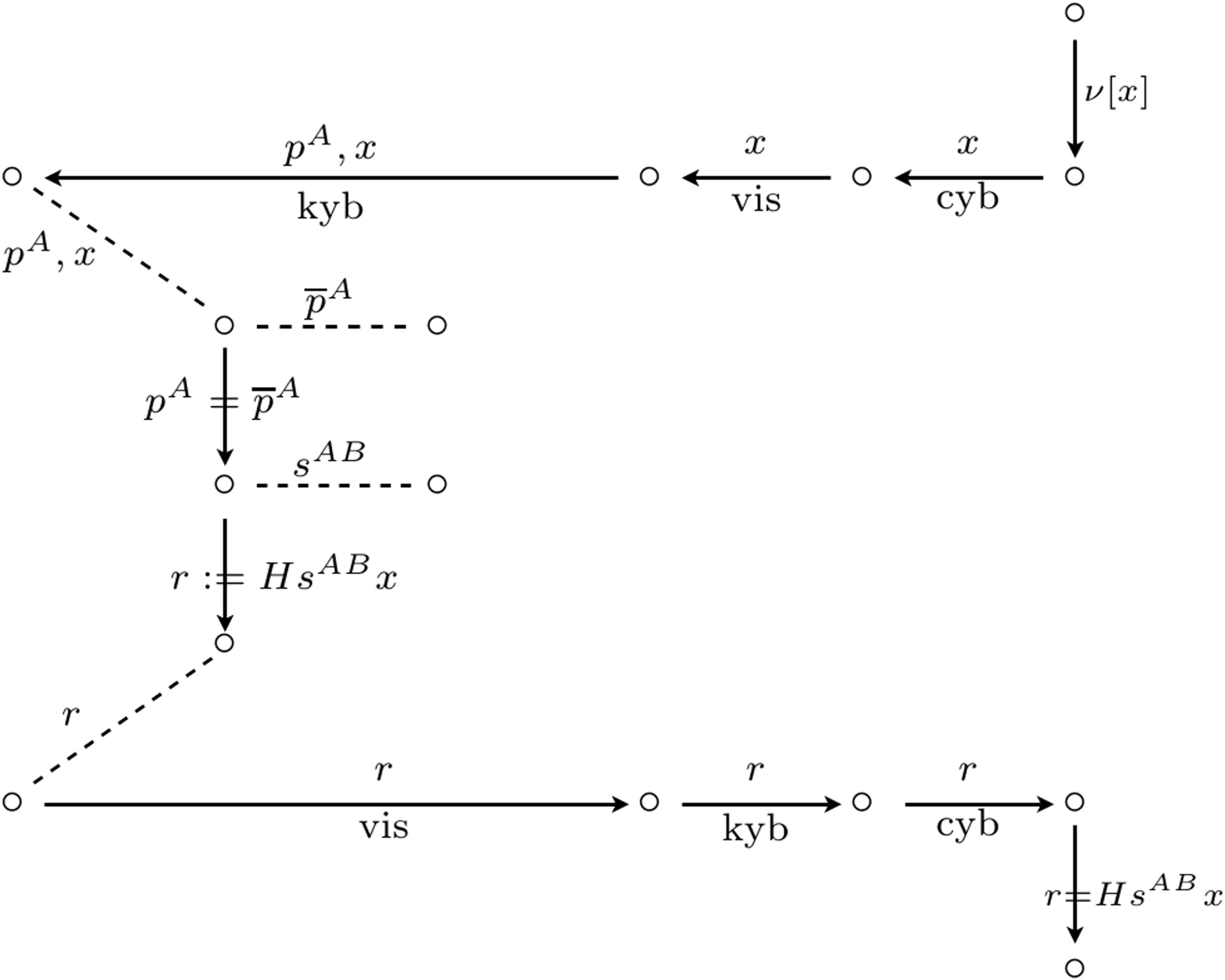}
  \end{center}
\vspace{-6ex}
\caption{The desired run of the CAP procedure.}
\label{fig_ex_CAP_run}
\vspace{-2ex}
\end{figure}

Based on this structure, runs are drawn (usually one run, the desired/secure run). The Figure~\ref{fig_ex_CAP_run} (taken from \cite[Fig.7]{pavlovic11ActorNet_journal}) graphically represents the desired run for the CAP authentication. For a better visual association, the structure of the configurations is displayed at the top, in a more simplistic manner.

The run in Figure~\ref{fig_ex_CAP_run} shows several aspects of ANPs. Internal actions are drawn vertically, whereas communication between configurations are drawn horizontally. The actions are displayed on the arrows, as well as the channel type that is used for communication. There are actions of generation of fresh content, of transmission of information, of polyadic communication (tuples of messages are sent), tests, and assignments. The dashed lines again are related to configurations made from combination of other configurations, but in a run represent sharing of information.

In this run the Bank generates a fresh value $x$ and sends it to the computer of Alice which communicates it to the human through a visual display channel. The human forwards this value and a password to the card reader through a keyboard input channel. The card reader and the card form a configuration inside which the information sent by the human is shared. This configuration compares the password send by the human to the one stored on the card $S_{A}$. If matched then the card gives away a long secret $s^{AB}$, which is difficult for a human to remember, opposed to a password. The configuration $Q$ hashes this secret and the fresh value into a response which the card reader displays to the human to copy and forward it through the computer of Alice to the Bank's computer. The long secret $s^{AB}$ is shared a priori between the card and the Bank. Therefore the Bank can perform the same calculation to generate a hashed value from this secret and the fresh value and to test it against the received response.

The fresh value is used to ensure that phishing cannot be done through recording just one session. The secret stored on the card is assumed to be a strong secret. The password is used only to make sure that the right human has made the configuration $Q$ by inserting her card into the card reader. Sending this weak password is done through a physical channel, which is assumed to be harder to attack.

\subsection{Encoding}\label{subsec_encoding}

Actor network procedures essentially consist of a structure of configurations together with events/actions causally ordered in the concurrent runs of the procedure/protocol. Configurations are partially ordered by a containment relation, which specifies which sub-configurations form a larger configuration. Configurations have attached channel ends (input and output ends). There is information flow inside configurations; more details will come later in the text.

We therefore, identify one nominal datatype built over a set of \textit{configuration names} \confNames. This datatype captures the partial order on the configurations. The terms that we will use are lists of configuration names describing reachability paths based on the parent-child relation between configurations; i.e., each configuration comes with the list of its ancestors.
\[
[l_{1},\cdots,l_{n}] \in \psiTerms\mbox{\ \ for\ \ }l_{i}\in\confNames.
\]
The list terms may also contain variables, and the names in the list may also be hidden behind a name binding restriction operation $\psiNew{\cdot}{}$. 

Channels and channel types do not bare much distinction in \cite{pavlovic11ActorNet_journal}. In consequence we will treat them in this paper as one and the same \textit{channel name}. If proper channel types are needed (like stating what kind of messages are allowed to be communicated) we could turn to the work on typed psi-calculi of \cite{Huttel11typed_psi} which extends the classic works on typed pi-calculus \cite{PierceS96typed_pi,sangiorgiwalker02pibook} or on the distributed pi-calculus \cite{dpi02,hennessy07dpi_book} where types capture resources. More complex types of channels, like the ones that \cite{pavlovic11ActorNet_journal} talks about, could be captured with complex processes that are processing the messages received, before forwarding them to the intended recipient. This is how channels like a \textit{random noise binary channel} would be described, or a \textit{lossy channel}.
Moreover, we do not restrict the formalism and do not assume that only one channel of one type exists between two configurations, as is done in \cite{pavlovic11ActorNet_journal}.

In consequence, the nominal datatype from before is enriched with a set of \textit{channel names} \chanNames\ which are paired with the list terms. This describes how a channel name is attached to the configuration described by the list term.
In consequence a channel is represented by a term:
\[
[l_{1},\cdots,l_{n}]c \in \psiTerms\mbox{\ \ for\ \ }c\in\chanNames,
\]
where the channel name is $c$ and the list $[l_{1},\cdots,l_{n}]$ denotes the configuration (and it's ancestor configurations) to which $c$ is attached.

Communication in psi-calculus on such a channel is defined with the psi-calculus process syntax from the previous section, e.g.:
\[
\psiOutput{[l_{1},\cdots,l_{n}]c}{N}\mbox{\hspace{2ex}in ANP notation would be\hspace{2ex}}\psiOutput{c}{[N]_{[l_{1},\cdots,l_{n}]}}
\]
where $N\in\psiTerms$ is usually a message term. Intuitively, this says that there is an output (sending) of the message $N$ on the channel $c$ in the configuration $[l_{1},\cdots,l_{n}]$.

We allow message terms to be constructed from some arbitrary signature, as it is allowed in the actor network procedures, and supported by the psi-calculi framework. For the purposes of this paper, the signature for building messages $N$ is left open, and is unimportant for the developments that we do further. For a specific application to a security ceremony or protocol, the designer decides on the message terms needed in the procedure. In this way we allow one to specify the minimal signature for message terms as needed for the specific protocol. Therefore the computation needed to analyze the specific protocol is dependent on the messages being sent part of the protocol.

One more nominal set \identNames\ keeps track of the \textit{identities} involved in the ceremony.
Identities are associated to configurations through a partial function, describing which identity controls which configuration; some configurations may be uncontrolled (thus the partiality of the function).
We use a pairing construct to form configurations which are controlled by some principal:
\[
i[l_{1},\cdots,l_{n}]\in\psiTerms\mbox{\ for\ } i\in \identNames.
\]

Channels can be attached to both uncontrolled configurations, like we did before, as well as to controlled configurations:
\[
i[l_{1},\cdots,l_{n}]c\in\psiTerms\mbox{\ for\ } i\in \identNames,l_{i}\in\chanNames,c\in\confNames.
\]
This example of term involves all three nominal sets, and is arguably the most complex form of terms we will use for communications in an ANP. More complication would come only from the specific term algebra of the messages that a designer chooses for a specific protocol.

In the actor network procedures, the communication operations which are not controlled, i.e.,
\[
\psiOutput{[l_{1},\cdots,l_{n}]c}{N}\mbox{\hspace{1ex} as Output, or\hspace{4ex} }\psiInput{[l_{1},\cdots,l_{n}]c}{\psiNew{\tilde{a}}{N}}\mbox{\hspace{1ex} as Input}
\]
are called \textit{events}, whereas the controlled ones, like 
\[
\psiOutput{i[l_{1},\cdots,l_{n}]c}{N}\mbox{\hspace{3ex}or\hspace{4ex} }\psiInput{i[l_{1},\cdots,l_{n}]c}{\psiNew{\tilde{a}}{N}},
\]
are called \textit{actions}. But there is no essential difference, and thus the graphical notation does not make a distinction between the two.

The structure of the actor network procedure is intended to capture how information is shared within the sub-configurations. In particular, information could flow from a main configuration to its sub-configurations when received on a channel attached to the main configuration. Opposite, information could flow from a sub-configuration to its parents (and ancestors) configurations, and be sent out through channels attached to an ancestor, and not to the originator configuration.
This form of information sharing is rather open and liberal in ANPs, which would leave room for a lot of hidden flow of information.

This means that there is not much control on the sharing of information in ANPs. It may be that the parent configuration wants to share some information only with part of its sub-configurations.
In consequence, here we allow for more control on the information flow so a communication action can specify explicitly to which sub-configuration it wants to communicate.
Moreover, we allow for hiding of the internal structure of configurations. When hiding is used then sharing can be implemented only internally, by first communicating with the main (public) configuration, which in turn decides to which sub-configurations to forward the message (possibly changed).

Until now we have encoded the communication structure of an actor network procedure through the terms \psiTerms\ of psi-calculus. The approach that we took above is inspired by the nested distributed pi-calculus of \cite{huttel13private}, where locations can be nested (i.e., a location can have sub-locations). Depending on the list terms that one defines, different relations between the configurations can be obtained. For the ANPs in particular, we are aiming for a partial order relation.

But the formalism for ANPs should subsume existing formalisms for security protocols and communicating processes. Indeed, if the lists defined above are always empty, and only channel names are used, then we obtain the communication mechanism of pi-calculus. Assume that no identities are present in the terms. The distributed pi-calculus is then obtained when working only with singleton lists, i.e., a flat structure of the locations (or configurations). Ambient calculi \cite{cardelli98mobileambients} or bigraphs \cite{milner09book} are obtained by making a tree-like structure between the configurations.

Internal sharing can be captured using local communication, hidden from outsiders.
The same approach we use to model local actions, like assignment or generation of fresh names as in the CAP example. Sending a fresh value on the cyber channel in the CAP example would be encoded as:
\[
\psiNew{\mathit{fr}}{\,\psiOutput{B[l_{C}]cyb}{\mathit{fr}}.\psiEmptyProc}
\]
For ease of notation, henceforth we will not add the empty process at the end.
To encode a local/internal action we use a private channel on which the configuration communicates with itself, as:
\[
\psiNew{\mathit{loc}}{(\psiOutput{B[l_{C}]loc}{\mathit{fr}} \psiPar \psiInput{B[l_{C}]loc}{(\lambda x)x})}.
\]

\begin{example}
The local generation of the fresh value $\mathit{fr}$ by Bank's computer $B[l_{C}]$, which is then communicated on the cyber channel is thus modeled as:
\vspace{-1ex}\[
\psiNew{\mathit{loc}}{(\psiNew{\mathit{fr}}{\psiOutput{B[l_{C}]\mathit{loc}}{\mathit{fr}}} \psiPar \psiInput{B[l_{C}]\mathit{loc}}{(\lambda x)x}.\psiOutput{B[l_{C}]\mathit{cyb}}{x})}.
\vspace{-1ex}
\]
The restriction operator $\nu$ applied to the name $\mathit{loc}$ makes this communication channel visible only inside the scope of $\nu$, whereas the restriction of the name $\mathit{fr}$ models the uniqueness, thus freshness.
\end{example}

Up to now we have made use only of the \psiTerms\ nominal datatype of the psi-calculi, not needing the logical part offered by the assertions \psiAssertions\ and conditions \psiConditions. The terms we described until now, part of \psiTerms, are used to capture the complex communication structure of ANP. This was used in conjunction with the process syntax for input/output, parallel composition and name restriction. 

The \textbf{case} process of psi-calculus is powerful, offering both non-determinism as well as conditionals.
Actor network procedures do not involve non-determinism, the same as the spi- and applied pi-calculus. These require only conditional constructs. This is natural when thinking that these are formalisms for describing security/communication protocols, i.e., deterministic runs of such protocols. But having the non-deterministic choice possibility in psi-calculi opens up for more modeling opportunities, like when wanting to refine the model of the human into a probabilistic one, involving probabilistic choices.

We have seen the necessity for the conditional in the CAP example, where terms were tested for equality. 
In consequence, we will include as conditions in \psiConditions, tests for equality for any two nominal terms from \psiTerms:
\vspace{-1ex}\[
M=N \in \psiConditions \mbox{\ \ for any\ \ }M,N\in\psiTerms.
\]
The entailment relation defines when two terms are equal wrt.\ some assertion:
\[
\psiassertion\psiEntailment M=N \mbox{\ \ iff\ \ } \vdash_{\Sigma}M=N
\vspace{-1ex}
\]
where $\Sigma$ is the signature over which the message terms are built, and $\vdash_{\Sigma}$ is the equational logic entailment relation wrt.\ the signature $\Sigma$. In other words, $M=N$ is decided only looking at the terms, using syntactic unification in the term algebra described by the signature $\Sigma$. In many cases equations are defined for such a term algebra, which need to be considered when deciding the equality of two terms. This would then involve working modulo these equations; we use the same notation $\vdash_{\Sigma}$ for the case when equations are part of the definition of the algebra of terms too.

Testing for equality of message terms does not depend on the assertions. 
Nevertheless, we will use assertions to model the partially ordered runs that actor network procedures use. This way of capturing true concurrency models (like the pomsets \cite{gischer84PhDpomsets,Pratt86pomsets} used by ANPs) in psi-calculi is inspired by the recent \cite{hakon13nwpt}.
Actor network procedures is among the few formalisms that acknowledges the need for true concurrency \cite{winskel95modelsCategory} when modeling protocols, instead of interleaving concurrency or linear runs as most process algebras approaches do. Actor network procedures describe a run to be a \textit{pomset}, i.e., a partially ordered multiset. Knowing that a linear run (word, string) is a totally ordered multiset (because symbols may appear multiple times in the string), then a pomset is a partially ordered run, i.e., a run where some of the actions are concurrent, not all being linearly ordered. It is natural to model a run of a protocol or ceremony as a partially ordered run because there are several parties involved, often distributed, thus executing actions/communication in parallel, concurrently. A single participant may be deterministic and sequential, thus exhibiting a linear run. But put together several participants exhibit a partially ordered run of the whole ceremony. More order constraints can come from the communications, where e.g., input actions depend on (i.e., must come after) the respective output actions.

Actor network procedures are declarative when defining their runs, in the same style as true concurrency models like event structures \cite{Winskel86} or pomsets, or as languages are, as opposed to automata or process algebras which describe their runs in an operational manner.
Psi-calculi allow also a declarative style of defining dependencies between actions by using the logic captured in the entailment relation and the assertions. In this way we have the full descriptive power of true concurrency models so to capture conjunctive (or disjunctive) dependencies as is the case with pomsets. This cannot be captured only through the sequence operation of the psi-processes.

We define assertions \psiAssertions\ to contain sets of communication terms, e.g.:
\[
\{\psiInput{i[l_{1},\dots,l_{k}]c}{N},\psiOutput{i[l_{1},\dots,l_{k}]c}{N},\dots\}\in\psiAssertions\mbox{\ \ with\ \ }i[l_{1},\dots,l_{k}]c\in\psiTerms\mbox{\ and\ }N\in\psiTerms.
\]
In consequence, conditions are also enriched with such actions by combining them with conjunction. The entailment is defined to treat conjunction appropriately, as in classical logics. For the new kind of conditions the entailment is just set-containment:
\[
\psiassertion\psiEntailment\psiInput{i[l_{1},\dots,l_{k}]c}{N} \mbox{\ \ iff\ \ }\psiInput{i[l_{1},\dots,l_{k}]c}{N}\in\psiassertion.
\]

With the conditions and assertions in place we can capture orders on the communication actions in the form of pomsets as follows. Each action is conditioned by a set of other actions on which it depends. In this way the action cannot be performed until the condition is met, i.e., all the actions on which it depends have been done. Thus,
\[
\psiCase{\varphi:\psiInput{i[l_{1},l_{2}]c}{N}.P}\mbox{\ with\ }\varphi=\{\psiOutput{j[l_{1}]d}{N'},\psiInput{i[l_{1},l_{2}]b}{N''}\}
\]
describes the fact that action $\psiInput{i[l_{1},l_{2}]c}{N}$ must come after the two actions from the condition have been executed.
The knowledge that an action has been executed is gathered in the context assertion through assertion processes which are left behind after each execution of an action. This is easily done by changing the continuation of an action:
\[
\psiInput{i[l_{1},l_{2}]c}{N}.P \mbox{\ \ becomes\ \ }\psiInput{i[l_{1},l_{2}]c}{N}.(P\psiPar\psiAssertOp{\psiInput{i[l_{1},l_{2}]c}{N}}).
\]
After the action is performed, the trailing assertion will become available to both the continuation and the other parallel processes, as part of the context assertion collected by the frame of the parallel process.

The actor network procedure formalism uses the PDL (Procedure Description Logic) to reason about runs. PDL\footnote{PDL for actor network procedures should not be confused with Propositional Dynamic Logic \cite{harel00dynamicLogic} (usually abbreviated PDL, or DL for the higher order case) used for reasoning about programs. On the other hand, propositional dynamic logic is a modal logic that reasons about actions in general, and could also be used for reasoning about ANPs once the special basic formulas and actions are set, as done for the ANPs. In fact, the reasoning style of PDL for ANPs resembles much the past temporal logic style of reasoning, and temporal reasoning can be done with propositional dynamic logic too.} uses two kinds of basic formulas: one states that an action has been executed; and another states that some action has been executed before another action.
Above, the assertions capture only the first kind of PDL basic formula.
We now add another assertion that stands for the second kind of PDL formula. This second kind of assertions capture a whole pomset in the assertions only. This pomset is available to the process for inspection, during the execution, and it captures the partial run so far. It is like a history in Hoare-style reasoning, only that in our case it is a partially ordered history.

We thus add to the assertion terms, dependency pairs of actions, giving the possibility to describe partial orders of actions as assertions:
\vspace{-1ex}\[
\psiInput{i[l_{1},\dots,l_{k}]c}{N}\dependsPomset \psiOutput{j[l_{1},\dots,l_{k}]d}{N'} \in \psiAssertions
\vspace{-1ex}
\]
signifying that the right-hand action depends on the left-hand action. If, moreover, both these actions are part of the assertions set then we conclude that the left action happened before the right action.

A question to ask is how do such dependency pairs get into the assertion set. Trailing assertion processes would be used, the same as we did earlier, only that more care needs to be taken when defining the assertion composition operation. We are not just using set union, but for building dependency pairs we must also achieve the transitivity property of the partial order relation we want to maintain.

\begin{example}
For the CAP run in Figure~\ref{fig_ex_CAP_run} the execution of the process would reach a point (e.g., upper left-most corner) where the environment assertion $\psiassertion$ would contain a dependency pair 
\vspace{-1ex}\[
\psiOutput{A[l_{C}]\mathit{cyb}}{\mathit{fr}} \dependsPomset \psiInput{A[l_{I}]\mathit{kyb}}{(p^{A},\mathit{fr})},
\vspace{-1ex}
\]
saying that the action of Alice of typing at the keyboard of the password and the fresh value is dependent, thus should come after, the computer of Alice having received the fresh value.
\end{example}

We have thus covered all aspects of the actor network procedures graphical formalism through the psi-calculus operational formalism. The structure of the configurations has been captured through the nominal datatypes, and the message terms have been treated the same as in ANPs. The definition of the pomset runs of a ANP was done by making use of the encoding of the dependencies between the actions using the assertions and conditions. Communication is done through channels attached to configurations, and internal actions are modeled also as communications but on private channels.

We have thus seen use of all psi-constructs for building processes: input/output used for communication and simulation of other actions like assignment; the \psiCase{} \!for modeling conditionals (and Hoare-style pre-conditions); restriction of names for modeling private communications and fresh values; parallel composition for putting several identities in the protocol to run together; and assertion processes for capturing the Hoare-style guarantees. The \textit{replication} construct has not been used; but it is essentially useful when needing to model ceremonies that can run through several sessions.

\section{Conclusions and Outlook}

There are many possibilities that psi-calculi open up for modeling ubiquitous systems and security ceremonies, where humans are part of the system and are intended to be modeled in a more faithful way.
The work we undertook in this paper shows that the psi-calculi framework is expressive enough to capture faithfully complex formalisms like the actor network procedures, and even with some generalizations thereof. We could easily capture concurrent computations in psi-calculi, besides sequential ones. 

The logic that psi-calculi offers is opened to be tailored to the application needs. In the case of ANPs we could use it to capture the Hoare-style reasoning that ANPs use. Such a reasoning is essential for security ceremonies where the assumptions should be made explicit, opposed to what usually is done with formalisms for security protocols where many assumptions are left underspecified.

The term construction mechanisms are also rather liberal in psi-calculi. This offers the possibility to define with any degree of detail needed, the message terms exchanged in the ceremony. At the same time we are not constrained when defining communication terms. This allows for modeling a great wealth of communication mechanisms. We have exploited this second freedom in the construction of the nominal terms representing the complex communication structure of ANPs.

For ceremonies in particular, we are interested in more faithful modeling of the human nodes where we would like to either leave room for errors, i.e., using non-determinism, or we would like to integrate a statistical model of the human, using probabilistic choices. These go beyond what current actor network procedures allow. But psi-calculi can accommodate probabilistic models, e.g., by going through CC-pi \cite{BuscemiM07ccpi,NicolaFMPT05qosCCpi} which has already been treated as a psi-calculus. But the work on probabilities and psi-calculus is not yet mature and still needs more investigation.

An interesting future direction for a graphical formalism line the ANPs for modeling ceremonies is the notion of action refinement \cite{GlabbeekG01refinement}. This is a technique for building (and working with) models in an incremental way, starting from an abstraction and refining actions into more concrete implementations.
Action refinement is well behaved for true concurrency models and their equivalences like history preserving bisimulation.
But it is not studied how to do action refined for graphical languages (except for the statecharts \cite{Harel87statecharts}), and not for ANPs either.
Refinement for ANPs would allow a ceremony designer to refine abstract models, both the configuration structures and the runs. By refining, one can expand single actions into more complex runs, or can expand one configuration with sub-configurations.

Action refinement is a technique for building models compositionally, in a top-down manner, whereas process algebras have a compositional approach where they build a model from components plugged together using operators like choice, sequential, or parallel composition.
Action refinement can work well in combination with the compositional approach of process algebras; and we encourage this combination.

The structure of the configurations in an ANP is static. It is not the case that during the execution of the ceremony some configurations are broken, disappear, or loose some of their sub-configurations.
But in the psi-calculi one can easily model a dynamic structure, similar to how the ambients are dynamic in ambient calculus \cite{cardelli98mobileambients}, or how communication changes locations in distributed pi-calculus \cite{hennessy07dpi_book}. The recent bigraphs formalism \cite{milner09book,Grohmann08SecurityCryptAndDirectedBigraphs} is especially focusing on how the structure of the system changes; the execution is defined in terms of change of structure (and interaction links). We see great possibility for investigating change of structure in ANPs, starting from the encoding we have given in this paper, and from the investigations of the above formalisms wrt.\ the psi-calculi framework.

\bibliographystyle{eptcs}
\bibliography{bib}

\end{document}